\documentclass[aps,prb,twocolumn,groupedaddress,floatfix]{revtex4}
\usepackage{graphicx}

\newcommand{\LSCO}{La$_{2-x}$Sr$_{x}$CuO$_{4}$}
\newcommand{\U}[1]{\ensuremath{\,\mathrm{#1}}}
\newcommand{\Tc}{\ensuremath{T_\mathrm{c}}}

\newcommand{\vect}[1]{\ensuremath{\mathbf{#1}}}
\newcommand{\degr}{\ensuremath{^\circ}}

\newcommand{\eulerbold}[1]{{\usefont{U}{eur}{b}{n}#1}}

\newcommand{\utaub}{\mbox{\eulerbold{\char28}}}

\begin{document}


\title[Automation using LabVIEW]
      {Automated operation of a home made torque magnetometer using LabVIEW}


\author{Stefan Kohout}
\email[current e-mail address: ]{kohout@physik.unizh.ch}
\author{Joseph Roos}
\author{Hugo Keller}
\affiliation{Physik-Institut, Universit\"at Z\"urich, 
Winterthurerstrasse 190, 8057 Z\"urich, Switzerland}


\date{\today}

\begin{abstract}

In order to simplify and optimize the operation of our home made torque
magnetometer we created a new software system. The architecture is
based on parallel, independently
running instrument handlers communicating with a main control program.
All programs are designed as command driven state machines which
greatly simplifies their maintenance and expansion. Moreover, as
the main program may receive commands not only from the user
interface, but also from other parallel running programs, an easy
way of automation is achieved. A program working through a text
file containing a sequence of commands and sending them to the
main program suffices to automatically have the system conduct
a complex set of measurements. In this paper we describe the
system's architecture and its implementation in LabVIEW.

\end{abstract}


\maketitle

\section{Introduction}
In modern condensed matter research most interesting subjects are only
subtle effects which can be investigated only by thorough and
systematic studies of large numbers of samples. Even though first
investigations have to be done by hand, a lot of time can be saved with
automated measurement setups. Such automated systems are already
widely used in large scale experiments, but most small laboratory
experiments, even though computer controlled, do not allow for
automated measurements. Automation is present, to some extent,
in order to facilitate measurements in that there frequently are
possibilities to have the system execute a certain measurement
automatically, but covering easily large parameter spaces is
often not possible. Commercially available complete measurement
systems, on the other hand, seldom come with sophisticated control
software without possibilities for programming long measurement
sequences. Of course such software systems are the result of expensive
software development which is beyond the possibilities of an
ordinary research laboratory. Even though there are commercial
programs available for some of the instruments that constitute an
experimental setup, the important part is the interplay between
them. Consequently most control software is written by the
scientists themselves, who face the lack of time, money and
manpower to develop extensive automation software.

In this paper we present an easy way
of creating control software which offers possibilities of programming
complex sequences and automatically executes them\cite{sourcecode}.
This is shown
to be achieved with moderate development effort using a common
laboratory programming language. We will first present the different
architecture approach needed to achieve this goal, after which the
addition of automation is a small step.

\section{software for laboratory equipment}
Programs created for control of experiments need to perform several
tasks. Firstly, they have to be able to send control commands to the
instruments and receive the measured data. Secondly, these data are
to be processed and displayed and eventually user input needs to be
translated to control commands. 
Various development platforms offer vast libraries of procedures
to interface instruments, create user interfaces and perform
complicated data processing. These help to reduce the workload
associated with creating such software. LabVIEW\texttrademark \cite{labview},
a development environment from National Instruments\texttrademark\
for creating programs (called virtual instruments or shortly VIs) in
its own graphical programming language ``G'', is probably best
known and most widely used for such applications. ``G'' offers all the
flow control structures like loops and conditional branches found in
any other programming language. Moreover, any VI can easily be used in
any other VI as a subVI.
LabVIEW VIs consist of a user interface (UI) and a block
diagram (BD) containing the actual code. Programming is done by modelling
data flow, where graphical representations of functions and procedures
are interconnected by lines, usually called wires. The designation VI
stems from the similarity of such a program to an actual instrument,
the UI obviously corresponding to the instrument's front panel and the
BD to its internal wiring.

A usual way of creating LabVIEW software for measurement
control is by writing a main VI containing the UI and the logic for
acting appropriately on the user input as well as processing,
displaying and saving the data. Communication with the instruments is 
performed by driver subVIs which are regularly executed by the main VI.
When such a driver VI is called to perform a query on an instrument
it sends the necessary command to the instrument, waits some time for 
the instrument to prepare the answer and finally reads this response
from the instrument. Usually this process takes tens to hundreds of
milliseconds. Assuming the whole measurement setup consists of several
instruments, the main VI may be organised in two different ways.
Either all driver VIs are called sequentially, causing the time needed
to collect all data to grow with the number of instruments. Another
apprach would be to call the driver VIs in parallel, which is possible
thanks to the inherently multithreading architecture of LabVIEW. In
this case, however, all drivers would attempt to access the
instruments at the same time. This would result in a ``traffic jam''
in case the instruments are connected to a single interface bus. Some
drivers would be forced to wait until the others have finished their
writing to the bus. Moreover, as some instruments take measurements
less often than others, many operations on the bus would be
unnecessary because no new data would be obtained.

In this paper we present the use of independent driver VIs, which we
call handlers, running in parallel
and communicating with a main VI by means offered by LabVIEW. This
allows for a more efficient use of the interface bus employed to connect
the instruments and results in a higher data acquisition rate.
Moreover, by employing a ``state machine'' (SM) architecture such
programs become easier to extend in functionality, to maintain and
most importantly allow for the control by a separate program and
consequently automation.

\section{experimental setup}
\begin{figure}
\center
\includegraphics[width=80mm]{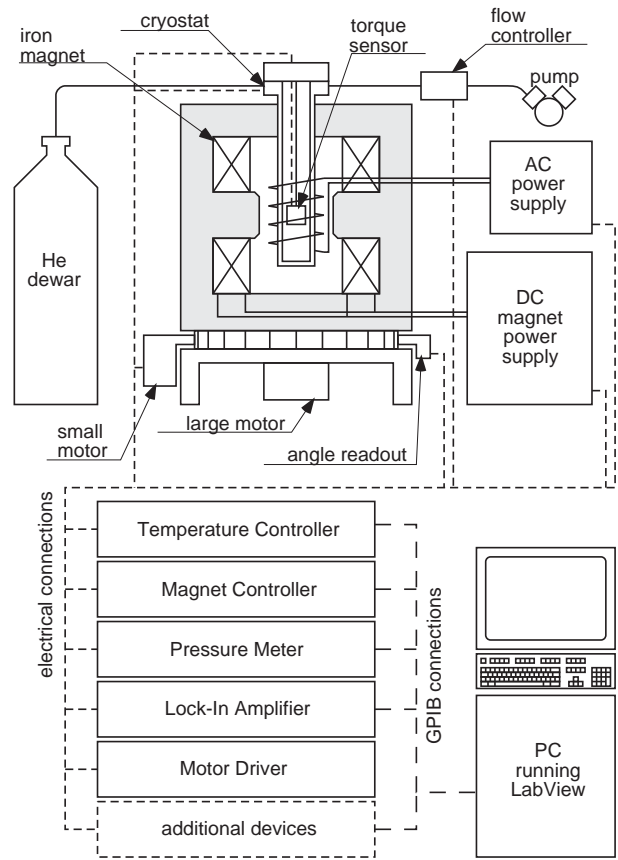}
\caption{\label{torquesetup}Torque measurement setup overview,
which was automated using the presented software. A cryostat is placed
between the poles of an iron yoke magnet, which is freely rotatable.
The torque sensor is inserted into the cryostat and connected to
readout electronics. All instruments needed to control the experiment's
state are connected to a personal computer.}
\end{figure}
The programs presented here were developed to control and automatise a
torque magnetometry apparatus which was built in our group\cite{Willemin1998a,Willemin1998b}.
Such a device is used to measure a sample's magnetic moment $\vect{m}$
by the torque 
\begin{equation}
\utaub = \mu_0 \vect{m} \times \vect{H}
\end{equation}
it experiences due to a magnetic field $\vect{H}$. It is well suited
for investigation of anisotropic magnetic phenomena as found in most
high temperature superconductors. Torque magnetometry is
complementary to most other magnetometry techniques in that it
is only sensitive to the part $m_\perp$ of \vect{m} perpendicular to
the applied field. A torque measurement
is fast --- one measurement taking a fraction of a second only ---
and due to the proportionality $\tau \propto H$ reaches high
sensitivities for $m_\perp$ in high fields. Our home made torque
magnetometer system, shown schematically in Fig.~\ref{torquesetup},
consists of a flow cryostat between the poles of an iron yoke magnet
which is sitting on a rotatable support.
The torque sensor with a sample mounted on it is inserted into the
cryostat and connected to a Lock-In Amplifier (LIA) for read out.
Details of the measurement principle are beyond the scope of this
article and are described elsewhere \cite{Willemin1998a,Willemin1998b}.
All devices needed to control and measure the system's state
are connected to a Windows PC via an IEEE-488 General Purpose
Interface Bus (GPIB), RS-232 serial connections and indirectly
via additional analog and digital input and output ports present
in the LIA instrument. The main parts are the EG\&G Model 7265 LIA,
a Lakeshore DRC 93A temperature controller, and the Bruker BH-15
magnetic field controller. Additional devices such as a
pressure transducer with read out electronics for monitoring
the exchange gas pressure in the cryostat or current sources
and volt meters for specialized applications may also be connected
via the GPIB. The GPIB is an interface bus which is widely used in
scientific instruments. It features 8-bit parallel data transfer,
handshaking and real-time response capabilities.

\section{software system architecture}

\begin{figure}
\center
\includegraphics{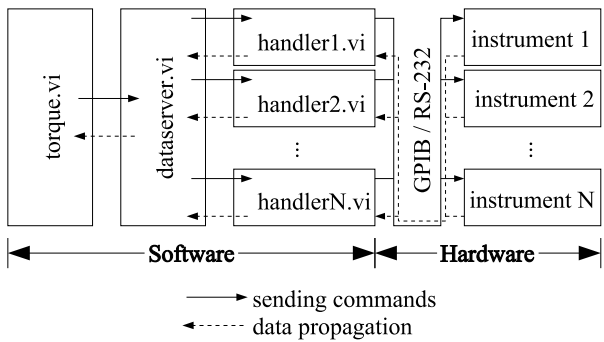}
\caption{\label{overview}Architecture of the torque control
software system. All VIs (torque.vi, dataserver.vi and the
handler*.vis) execute in parallel. Commands are sent along the solid
right pointing arrows and data propagates back along the dashed left
pointing arrows.}
\end{figure}

The architecture of the newly developed control software is
shown in Fig.~\ref{overview}. Each instrument connected
to the system is represented by a VI counterpart called \emph{handler.vi}.
All handlers are managed by the \emph{dataserver.vi} VI which
communicates with the \emph{torque.vi} VI, which is the main
application. All these VIs run independently in parallel.
This way each \emph{handler.vi} can be optimised to take best advantage
of the instrument it is built for. This includes the waiting times
needed for communication, an optimized data rate based on varying
needs as well as the use of each instruments ability to signal special
events via the GPIB. Since all \emph{handler.vis} run in parallel, their
individual write--wait--read cycles needed to talk to the instruments 
are interlaced, thus reducing the bus' idle time.
Moreover each instrument is talked to only when necessary thus
reducing the bus occupation while retaining data quality. This can be
optimised particularly well by exploiting the service request (SRQ)
functionality of the GPIB. Each instrument can signal a number of
events to the GPIB controller by asserting the special SRQ line. Such 
events might be error conditions but can also be indicators of data
availability. As an example the Lakeshore temperature controller is
programmed to assert the SRQ line whenever a new temperature reading
is ready. As this occurs only every two seconds, the instrument is
read only when really necessary instead of reading the same data
several times per second. Even instruments not offering such
functionality can be optimised by reducing the rate at which the
\emph{handler.vi} is instructed to read the instrument. This enables the 
more crucial measurements to be read more often resulting in data
taken at a higher rate and resulting in better quality.


Because the \emph{handler.vis} are not called as subVIs by the main
VI a special means of communication needs to be established.
Here we present the use of \emph{queues} for sending commands to
the \emph{handler.vis} and \emph{DataSockets} for receiving the
measured data. A \emph{queue} is a first--in--first--out style memory
construct which is offered by LabVIEW. It may contain a fixed or unlimited
number of string entries, in our case commands. By use
of special subVIs any VI can append commands to a \emph{queue's}
end or retrieve the oldest commands. Any read entry is automatically
removed. \emph{Queues} are identified by a name, making access
to them fairly easy. In most applications a given \emph{queue} is read
by only one VI whereas several VIs may write to it.
\emph{DataSockets} are memory constructs as well, identified
by a unique name, but only contain the most recent datum. Their
data type can be freely chosen among the data types in LabVIEW.
The \emph{DataSockets} used in our case are arrays of floating
point numbers containing a \emph{handler.vi's} main data.
The \emph{dataserver.vi} mentioned above serves as an intermediate
VI which collects all the \emph{handler.vi's} data and puts
all together in a separate \emph{DataSocket} which is then read
by the \emph{torque.vi} main VI. Thus the main VI needs no knowledge
about which data to obtain from which instrument.


\begin{figure}
\center
\includegraphics{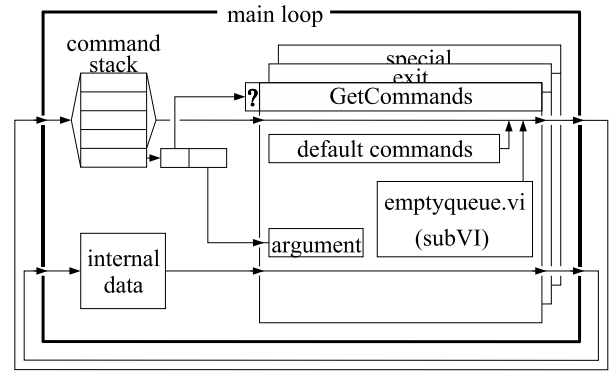}
\caption{\label{FIGparser}Schematic illustration of the VI's
basic structure. An all enclosing main loop executes infinitely.
The logic inside consists of a command
stack whose first element is divided into instruction and argument.
The instruction is used as the selector value into
a case structure containing the code for the individual
instructions. This results in the command parsing functionality needed
for the operation. Internal data needed for the VI's execution is
passed through each iteration and can be read and modified
by each command case.}
\end{figure}

In order for the VIs to be able to act accordingly
on the possible commands they must be given some command parsing
functionality. In fact such a command parser is every VI's
core part: Even the regular operations performed by the VIs are
put into commands which are executed repeatedly. Essentially, all
VIs are designed as command driven state machines (SM).
The use of the SM paradigm 
in LabVIEW programs was already proposed at several occasions
and given LabVIEW's capabilities this is not surprising.
Nevertheless, to our knowledge only few applications make use of this
architecture. The basic idea is that by being executed, a program goes
through various named states. The order in which these states are
visited may be fixed and defined in advance or the state to follow
might be determined based on the current state's result. The
implementation in LabVIEW is fairly simple and schematically shown
in Fig.~\ref{FIGparser}. An infinitely running loop
contains a case structure consisting of all the states. These
states are identified by character strings and are therefore
easily human readable. In contrast to other methods, where the
identification is by numbers or special enumeration data types,
this makes the structure easy to extend and maintain.
Additionally to these structures the VIs contain a command stack
and some internal data needed for execution. Upon startup, when
the command stack is empty, a default case (state) is executed.
Usually this is the ``GetCommands'' case. This case contains
the code needed to empty this VI's \emph{queue} and a set of default
commands which are put onto the command stack. When the main loop
is iterated for the second time, the oldest command is removed
from the stack, split into an instruction and optional
arguments, whereupon this instruction is fed into the case
structure selector, defining the case to be executed. This case
may add more commands to the stack or simply perform a specific
task. When the case is finished, the main loop iterates again,
the next command is removed from the stack and so on. Whenever the
stack becomes empty, the default case ``GetCommands'' is
executed again and refills it.

Because the \emph{handler.vis} are independent programs not having
to rely on being called regularly by a master VI they can be
used to carry out more complex tasks than just talking to
the instruments. As an example \emph{handlerLakeshore.vi}, the
\emph{handler.vi} for the Lakeshore temperature
controller contains logic to control the temperature by software
through control of the coolant flow in the cryostat. The flow
controller is connected to a separate digital--to--analog converter
(DAC), thus enabling the \emph{handlerLakeshore.vi} to control
it by sending commands to the DAC's \emph{handler.vi}
(\emph{handlerDAC.vi}).
Keeping track of the last few seconds of measured data,
calculating their time trends and publishing it to the \emph{DataSocket}
is coded into a command and performed by the \emph{handler.vis} as well.

\section{Automation}
\begin{figure}
\center
\includegraphics{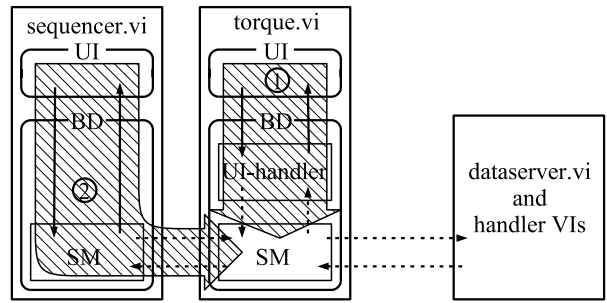}
\caption{\label{automation}All VIs consist of a User interface (UI) and
a block diagram (BD). In contrast to all other VIs the \emph{torque.vi's}
BD consists of the UI-handler part and the state machine (SM) itself,
both running in parallel. In normal, interactive operation of the torque
system, user actions on the \emph{torque.vi's} UI are translated by the
UI-handler into commands which are sent to the SM via a queue and then
propagate on to the dataserver and handler VIs (wide arrow (1)).
If an automated measurement is run, the \emph{sequencer.vi's} SM retrieves
commands from the text sequence on its UI, sends them via a queue to the
\emph{torque.vi's} SM from where they propagate on to the dataserver and
handler VIs (wide arrow (2)). The \emph{torque.vi's} SM sends confirmation
messages back to the \emph{sequencer.vi}. Solid black arrows
indicate direct access between the BD and the UI, whereas dotted arrows
represent data transmission via \emph{queues} and \emph{DataSockets}.}
\end{figure}

As mentioned earlier, all VIs are organized as state machines,
even the main VI \emph{torque.vi}. As shown in Fig.~\ref{automation}
every user action (button press, value change) on its user interface (UI)
is transformed into a command by the UI-handler which is then sent to
and processed in the SM. The SM then sends appropriate
commands to the \emph{dataserver.vi} and the \emph{handler.vis} (wide
arrow (1) in Fig.~\ref{automation}).
These two parts (UI-handler and SM) are independently running components
of \emph{torque.vi}. The communication between them is again ensured via
\emph{queues}. This enables other VIs, such as the \emph{sequencer.vi}
shown in Fig.~\ref{automation} to be used to control
the SM in \emph{torque.vi} programmatically by sending these commands
directly to the SM (wide arrow (2) in Fig.~\ref{automation}).

When automatic measurements are required, a sequence text file
is written containing the commands needed to accomplish
these measurements which is then read by the \emph{sequencer.vi}.
Additionally to the commands of \emph{torque.vi's} SM the
\emph{sequencer.vi} understands a set of flow control instructions
such as ``if'', ``while'' and ``for'' which are useful
for creating short sequences for repetitive tasks, as well as
the use of variables and their arithmetic manipulation and
comparison.

The \emph{sequencer.vi} parses through the sequence file by looking
for known keywords -- the commands. Any strings which are not
recognized as a keyword are treated as arguments to the preceding
keyword. The string \texttt{settemp 20 waittemp} present in a sequence
file would instruct the torque software to change the temperature
to 20\,K and wait for the cryostat to stabilize at this temperature.
In this example \texttt{settemp} and \texttt{waittemp} are keywords
and \texttt{20} is the argument to the keyword \texttt{settemp}.
Such sequencing possibilities are already well known in control
software of commercially available measurement equipment
(eg.\ SQUID magnetometers or the \emph{Quantum Design} Physical
Property Measurement System\cite{Quantum}).
Now such efficient and flexible data taking is also possible
with our home made torque magnetometer.

\section{Example of Application}
\begin{figure}
\center
\includegraphics{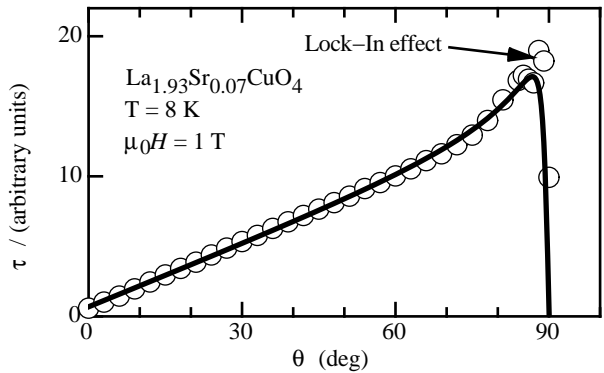}
\caption{\label{torque1}Angle dependent torque measurement (circles)
of an underdoped crystal of \LSCO\ with $x=0.07$ ($\Tc=17\U{K}$),
performed at $T=8\U{K}$ in a magnetic field $\mu_0H=1\U{T}$.
The solid line is a fit of a model derived by Kogan\cite{kogan}. The
deviation close to $\theta \approx 90\degr$ stems from the lock-in
transition.}
\end{figure}

In order to demonstrate the possibilities of such an automatable
measurement system we present some results of a systematic
study\cite{Kohout2005}
of the so called lock-in transition in the high temperature
superconductor \LSCO. Details about this effect can be obtained
from various other sources and are not discussed
here\cite{Blatter1994,Steinmeyer1994}.
Most easily this effect is visible in angle dependent torque
measurements and manifests itself as a deviation from an otherwise
smooth behaviour. An example of such a measurement is shown in
Fig.~\ref{torque1}, where the measured data points close to
90$^\circ$ deviate from a theoretical curve\cite{kogan} which fits
well to the remaining angle range.
The same model can also be used to describe data taken as a
function of magnetic field magnitude $H$ at a fixed angle.
It is commonly accepted that in first approximation the
magnetic moment $m=\tau/H$ of a superconductor is proportional
to $\ln(H)$.

Within our study we measured six \LSCO\ single microcrystals with
varying Sr content $0.07 \le x \le 0.23$ and critical temperatures \Tc\
varying from 17\U{K} to 35\U{K}. They were mounted on
a highly sensitive torque sensor and cooled below \Tc.
Field dependent measurements
($\mu_0H=0\ldots1.5\U{T}$ at $5\U{mT}$ steps with increasing and
decreasing field) were
taken at 60 field orientations ($\theta=-90\degr\ldots90\degr$ with
varying steps) and at about ten temperatures below the critical
temperature \Tc.
We emphasize that such extensive measurements would hardly be
possible without our software's automation possibilities. As each
field scan takes about six minutes, without automation user interaction
would be necessary at this interval during \emph{one week} to collect
all these data for one crystal. 
After writing the sequence and starting its execution, the
measurement system, on the other hand, finishes such a measurement
set within about \emph{three days} with no need of intervention.
The experiment is finished faster, because less time is lost
between consecutive field scans and because the measurement is
running day and night.
\begin{figure}
\center
\includegraphics{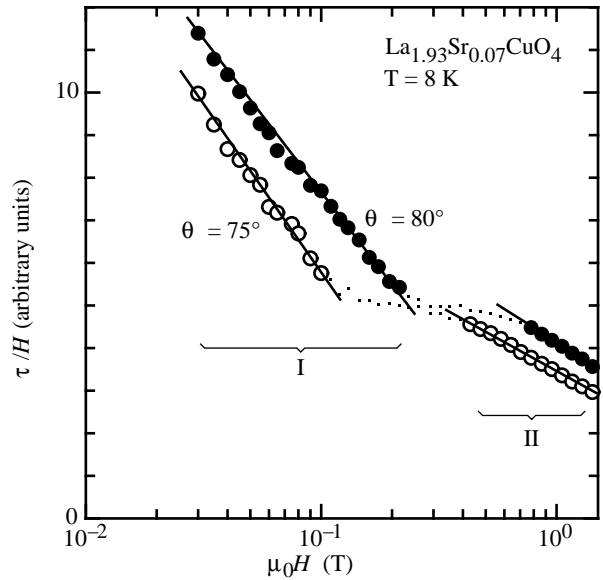}
\caption{\label{torque2}Field dependent measurement $\tau(H)$ of the
same \LSCO\ crystal as was used for the measurement in Fig.\ \ref{torque1}.
The angle of the magnetic field was fixed at $\theta=75\degr$ and $\theta=80\degr$. The measurements are plotted as $\tau/H$ vs.\ $\ln(H)$.
The lines are guides to the eye to show the two linear regions I
(low field) and II (high field).}
\end{figure}

\begin{figure}
\center
\includegraphics{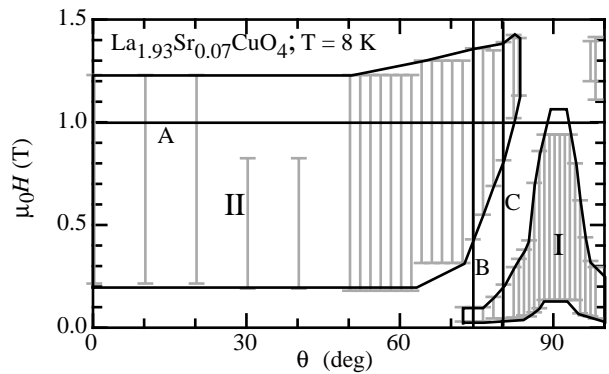}
\caption{\label{torque3}Summary of field dependent measurements
performed on a \LSCO\ single crystal at $T=8\U{K}$. Only the extents
of the linear regions such as shown in Fig.\ \ref{torque2} as a function
of field orientation $\theta$ are shown. The enhancement of the low-field
region I close to the ab-plane ($\theta \approx 90\degr$) is
clearly visible. The horizontal line A indicates the position
of the measurement shown in Fig.~\ref{torque1}. The
vertical lines B and C indicate the position of the measurements
shown in Fig.~\ref{torque2}.}
\end{figure}

We present here only one dataset of a single crystal taken at
one particular temperature. Such a dataset consists of 60
field scans taken at various orientations. The two field scans
shown in Fig.~\ref{torque2} illustrate the deviations of field
dependent data due to the lock-in transition. Clearly visible
are two regions (I and II) where $\tau/H$ is proportional to
$\ln(H)$. A comparison of these measurements to angle dependent
measurements at similar conditions indicate that region I corresponds
to the part where lock-in takes place, whereas data in
region II are well described by the theoretical curve
in Fig.~\ref{torque1}. By analysing the whole data set it is now
easy to investigate the evolution of these two regions as a
function of angle $\theta$. The result is shown in
Fig.~\ref{torque3}, where the extents of the two regions, obtained
from field dependent measurements, are plotted vs.\ the angle
$\theta$. The horizontal line A indicates the cut of the
measurement in Fig.~\ref{torque1} and the vertical lines B and
C the measurements shown in Fig.~\ref{torque2}.
The observed region, separating regions I and II manifests the
lock-in transition and can be understood in terms of a
model proposed by Feinberg and Villard\cite{Feinberg}.


\section{Acknowledgements}
This work was supported in part by the Swiss National Science
Foundation.

\newpage


\end{document}